\documentclass[aps,twocolumn]{revtex4}
\usepackage{amsfonts}
\usepackage{amsmath}
\usepackage{mathrsfs}
\usepackage{amssymb}
\usepackage{graphicx}
\usepackage{dcolumn}
\usepackage{bm}
\usepackage{color}
\usepackage[colorlinks,linkcolor=blue,anchorcolor=blue,citecolor=blue,urlcolor=black]{hyperref}

\newcommand{\ket}[1] {\left| #1 \right\rangle}

\begin{document}

\title{Superradiance-driven Phonon Laser}

\author{Yajing Jiang$^{1,\dagger}$, Hao L\"{u}$^{2,4,\dagger}$, and Hui Jing$^{3,}$}
\email{jinghui73@foxmail.com}
\affiliation{$^1$Department of Physics, Henan Normal University, Xinxiang 453007, China}
\affiliation{$^2$Key Laboratory for Quantum Optics, Shanghai Institute of Optics and Fine Mechanics, Chinese Academy of Sciences, Shanghai 201800, China}
\affiliation{$^3$Key Laboratory of Low-Dimensional Quantum Structures and Quantum Control of Ministry of Education, Department of Physics and Synergetic Innovation Center for Quantum Effects and Applications, Hunan Normal University, Changsha 410081, China}
\affiliation{$^4$University of Chinese Academy of Sciences, Beijing 100049, China}
\affiliation{$^\dagger$These authors contributed equally to this work.}


\begin{abstract}
We propose to enhance the generation of a phonon laser by exploiting optical superradiance. In our scheme, the optomechanical cavity contains a movable membrane, which supports a mechanical mode, and the superradiance cavity can generate the coherent collective light emissions by applying a transverse pump to an ultracold intracavity atomic gas. The superradiant emission turns out to be capable of enhancing the phonon laser performance. This indicates a new way to operate a phonon laser with the assistance of coherent atomic gases trapped in a cavity or lattice potentials.    	
\end{abstract}


\thispagestyle{empty}

\maketitle
\twocolumngrid

\maketitle

Rapid advances in cavity optomechanics (COM) based on coherent photon-phonon interactions have a variety of applications, such as quantum state transducers, highly-sensitive sensors, and quantum squeezing or chaos \cite{AKM2014,Metcalfe2014,Polzik2014,Wu2017,LXY2015,jingsr,jingsr2017,Hao17}.
As an acoustic analog of the optical laser, phonon lasers have been experimentally observed in e.g. cold ions \cite{Vahala2009}, electromechanical circuits \cite{Mahboob2013}, superlattices \cite{Kent2010} and COM devices \cite{Vahala2010}.
Quantum acoustic effects, such as two-mode correlations and sub-Poissonian distributions, have also been observed \cite{Lawall2014,Painter2015}. These pioneering works pave the way for the designs and applications of low-noise phonon devices \cite{Jing2014,Zhang2015,defect,Li2012rmp}.

Recently, superradiance or collective coherent emission from ensembles of emitters \cite{Dicke1954,cpl}, has been experimentally realized by pumping a collection of cold atoms \cite{Esslinger2010,Hemmerich2015,Roof2016}. Some exotic effects at the superradiant phase have also been observed, such as symmetric breaking \cite{Esslinger2011} and roton-type mode softening \cite{Esslinger2012}.
The successes in experiments have stimulated extensive theoretical works on e.g. superradiant solid \cite{Ji2013} and fermionic superradiance \cite{Keeling2014,Chen2014,Piazza2014,Zheng2016}. However, the possible role of superradiance in COM systems, especially amplifying sound, has rarely been studied.

We note that in previous works, superradiant emission has been confirmed to provide a driving field for single-cavity COM systems \cite{Santos2010,Liang2012}.
In these systems, the tunability of the atom-photon interactions provides a new degree of freedom to manipulate the mechanical modes. Here, we extend to study a phonon laser driven by superradiance. Our system consists of a cavity supporting a mechanical mode and another cavity containing an atomic Bose-Einstein condensate (BEC). The atoms are considered in one dimension along the cavity axis direction, which are pumped by a transversal field. The superradiance can emerge by increasing the pump power, leading to enhanced radiation pressure on the COM interactions and therefore, the enhanced phonon laser.

\begin{figure}[ht]
\centering
\includegraphics[width=7.7cm]{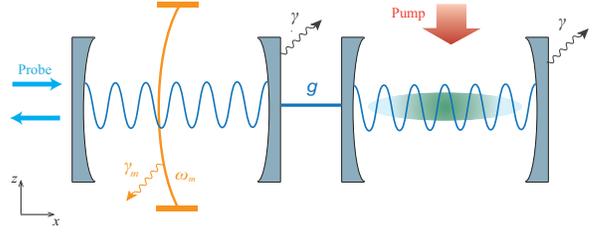}
\caption{Schematic illustration of the superradiance-driven phonon laser. The left cavity contains a movable membrane, and the right cavity is coupled to the left cavity with the strength $g$. A probe field with frequency $\omega_p$ is applied to the left cavity. A collection of two-level atoms are trapped in the right cavity, which is pumped by a transversal field at frequency $\omega_p$. The levels of atoms correspond to momentum states of the atoms.}
\label{fig1}
\end{figure}

As shown in Fig.\,\ref{fig1}, we consider two coupled cavities with the same resonance frequency $\omega_c$ and optical loss rate $\gamma$.
A weak probe light with frequency $\omega_p$ and wavelength $\lambda_p$ is applied to the left cavity, which contains a movable membrane. The membrane supports a mechanical mode with the frequency $\omega_m$ and damping rate $\gamma_m$, which is coupled with the light via COM interactions. A collection of $N$ two-level atoms (with atomic mass $m$) are trapped in the right cavity, which is pumped by a transversal field at frequency $\omega_p$
and Rabi frequency $\Omega_p=\mathcal{D} E_0/\hbar$. $\mathcal{D}$ is the transition dipole matrix element and $E_0$ is the electric field. The levels of an atom are denoted by the zero momentum state $\ket{0}$ and excited state $\ket{\pm k}$, respectively \cite{Esslinger2010,Esslinger2011}, where $k=2\pi/\lambda_p$ is the wave vector of the pump light. Then the collective operators of the atoms $\left\lbrace J_{\pm},J_{z}\right\rbrace $ are defined as
\begin{align}
J_{\pm}=J_x\pm iJ_y,~~~J_j=\sum_{i=1}^N\sigma_{j}^{(i)},(j=x,y,z),\nonumber
\end{align}
with the Pauli operator $\sigma_j^{(i)}$ for the $i$-th atom, which satisfy angular momentum commutation relations, i.e.,
$$[{J}_{\pm},{J}_z]=\mp{J}_{\pm},~~~[{J}_{+},{J}_{-}]=2{J_z}.$$
Here we note that $J_{\pm,z}$ correspond to the momentum states, not the internal states of the atoms.
This is reasonable for an atomic BEC, with a single collective internal state but different momentum states when disturbed by the external fields, according to the superradiance experiment with a cavity-trapped BEC \cite{Esslinger2010,Esslinger2011}.

The coupling of the atomic gas and the pump field in the right cavity can be described by the Dicke model \cite{Esslinger2010,Nagy2010}. In addition, the mechanical mode in the left cavity can be driven by the light tunneling between the two cavities. In a rotating frame at the pump frequency $\omega_p$, the Hamiltonian of this system can be written at the simplest level as
\begin{align}
{H}=&-\Delta_c {a}^{\dagger}_1{a}_1 - \Delta^{\prime}_c {a}^{\dagger}_2{a}_2+g({a}^{\dagger}_1{a}_2+{a}^{\dagger}_2{a}_1)\nonumber\\
&+\omega_m{b}^{\dagger}{b}-\chi{a}^{\dagger}_1{a}_1({b}^{\dagger}+{b})+\omega_r{J}_z\nonumber\\
&+\frac{\lambda}{\sqrt{N}}({J}_{+}+{J}_{-})({a}^{\dagger}_2+{a}_2),
\label{eq:Hd}
\end{align}
where $a_{1,2}$ or $b$ denote the annihilation operators of the optical modes or the mechanical mode.
$\Delta_c=\omega_p-\omega_c$ and $\Delta_a=\omega_p-\omega_a$ are the detunings of the pump light with respect to the cavity and atomic transition frequencies. In addition,
$$\lambda=\sqrt{N}\Omega_p g_0/\sqrt{2}\Delta_a,$$
$g_0$ is the atom-photon coupling strength, $\chi$ is the COM coupling rate, $\Delta^{\prime}_c=\Delta_c-NU_0/2$ with $U_0=g^2_0/\Delta_a$, and $\omega_r=\hbar k^2/2m$ is the recoil frequency.

The COM coupling typically much weaker than the atom-photon interaction, is assumed to have negligible effects on the superradiant phase transition. In the thermodynamic limit ($N\gg1$), the semiclassical equations of motion are written as \cite{Dimer2007}
\begin{align}
\dot{a}_1=&(i\Delta_c-\gamma)a_1 - iga_2,\nonumber\\
\dot{a}_2=&(i\Delta^{\prime}_c-\gamma)a_2 - iga_1 -\frac{i\lambda}{\sqrt{N}}(J_{+}+J_{-}),\nonumber\\
\dot{J}_{-}=&-i\omega_rJ_{-}+\frac{2i\lambda}{\sqrt{N}}J_z(a^{\dagger}_2+a_2),\nonumber\\
\dot{J}_z=&\frac{i\lambda}{\sqrt{N}}(J_{-}-J_{+})(a^{\dagger}_2+a_2),
\label{eq:dickemotion}
\end{align}
in which we have neglected the weak probe field.
The steady-state solutions of Eqs.\,(\ref{eq:dickemotion}) are
\begin{align}
a_{1,s}=&-\frac{iga_{2,s}}{\gamma-i\Delta_c},\nonumber\\
a_{2,s}=&-\frac{2\lambda(\Delta_c+i\gamma)J_{-,s}}{\sqrt{N}(u-iv)},\nonumber\\
J_{z,s}=&\frac{N\omega_r(u^2+v^2)}{8\lambda^2(\gamma v -\Delta_c u)},
\label{eq:steady}
\end{align}
with
$$
u=g^2+\gamma^2-\Delta_c\Delta^{\prime}_c,~~~
v=\gamma\left(\Delta_c+\Delta^{\prime}_c\right).
$$
We see that the steady-state values of the photons and the atomic polarization amplitudes depend on the optical tunneling rate between the two cavities.
In consideration of the pseudoangular momentum conservation
\begin{align}
|J_{-,s}|^2 + J^2_{z,s}=N^2/4,
\label{momentum}
\end{align} we obtain the critical phase transition point
\begin{equation}
\lambda_c=\frac{1}{2}\sqrt{\frac{\omega_r(u^2+v^2)}{|\gamma v-\Delta_c u|}}.
\end{equation}
When the two cavities are decoupled, i.e. $g=0$, the expression of the critical point reduces to that obtained in Ref.\,\cite{Esslinger2010}. If $\lambda<\lambda_c$, the system is in the normal phase, indicating that all the atoms are randomly-distributed, and the intercavity photon number is zero. Once $\lambda$ exceeds the critical point $\lambda_c$, the atoms are macroscopically
excited and the scattering from the pump to the cavity mode is greatly enhanced. The superradiant phase transition occurs, resulting in coherent collective light emissions.

In the following, we focus on the impact of the optical superradiance on the mechanical mode, using a semiclassical approach \cite{Vahala2010,Wang2014}. Equation\,(\ref{eq:Hd}) can be transformed into the optical supermode picture by defining the operators $a_{\pm}=(a_1\pm a_2)/\sqrt{2}$. By using the rotating-wave approximation, $2g+\omega_m,\omega_m\gg|2g-\omega_m|$, we have
\begin{align}
\mathcal{H}=&\mathcal{H}_0+\mathcal{H}_{\mathrm{int}},\nonumber\\
\mathcal{H}_0=&\omega_{+} a^{\dagger}_{+}a_{+} + \omega_{-} a^{\dagger}_{-}a_{-} -\frac{NU_0}{4}(a^{\dagger}_{+}a_{-}+a^{\dagger}_{-}a_{+}) \nonumber\\
&+\omega_m b^{\dagger}b + \omega_r J_z,\nonumber\\
\mathcal{H}_{\mathrm{int}}=&\frac{\lambda}{\sqrt{2N}}(J_{+}+J_{-})(a^{\dagger}_{+}+a_{+}-a^{\dagger}_{-}-a_{-})
\nonumber\\
&-\frac{\chi}{2} (a^{\dagger}_{+}a_{-}b +b^{\dagger}a^{\dagger}_{-}a_{+}),
\end{align}
where
$\omega_{\pm}=-\Delta_c\pm g+NU_0/4.$

The Heisenberg equations of motion of the system can then be written as
\begin{align}
\dot{a}_{+}=& \frac{i(NU_0+2\chi b)}{4} a_{-}-(i\omega_{+}+\gamma)a_{+} -\frac{i\lambda(J_{+}+J_{-})}{\sqrt{2N}},\nonumber\\
\dot{a}_{-}=& \frac{i(NU_0+2\chi b^{\dagger})}{4} a_{+} -(i\omega_{-}+\gamma)a_{-} +\frac{i\lambda(J_{+}+J_{-})}{\sqrt{2N}},\nonumber\\
\dot{b}=&(-i\omega_{m} - \gamma_{m})b + \frac{i\chi}{2} a^{\dagger}_{-}a_{+}\nonumber,\\
\dot{J}_{-}=&-i\omega_rJ_{-}+i\lambda\sqrt{\frac{2}{N}}(a^{\dagger}_{+}+a_{+}-a^{\dagger}_{-}-a_{-})J_z,\nonumber\\
\dot{J}_z=&\frac{i\lambda}{\sqrt{2N}}(a^{\dagger}_{+}+a_{+}-a^{\dagger}_{-}-a_{-})(J_{-}-J_{+}).
\end{align}

Then we define the population inversion operators as $p = a^{\dagger}_{-}a_{+}$  and $\delta n =a^{\dagger}_{+}a_{+} - a^{\dagger}_{-}a_{-}$, respectively. The dynamical equation of $p$ is
\begin{align}
\dot{p}=&(-2ig-2\gamma)p -\left(\frac{i\chi b}{2} + \frac{iNU_0}{4}\right)\delta n   \nonumber\\
&-i\lambda J_{-}\sqrt{\frac{2}{N}}(a_{+}+a^{\dagger}_{-}).
\end{align}
To obtain the mechanical gain, we set $\partial p/\partial t = 0$, $\partial J_{-}/\partial t = 0$, and
$\partial a_{\pm}/\partial t = 0$, with $\gamma\gg\gamma_m$ \cite{Vahala2010,Wang2014}, which leads to the following results
\begin{align}
J_{-}=&\frac{N\lambda (a^{\dagger}_{+}+a_{+}-a^{\dagger}_{-}-a_{-})}{\sqrt{4\lambda^2(a^{\dagger}_{+}+a_{+}-a^{\dagger}_{-}-a_{-})^2+2N\omega^2_r}},\nonumber\\
J_{z}=&\frac{N\sqrt{N}\omega_r}{2\sqrt{2\lambda^2(a^{\dagger}_{+}+a_{+}-a^{\dagger}_{-}-a_{-})^2+N\omega^2_r}},\nonumber\\
a_{+}=&-\frac{\lambda J_{-}(2i\gamma+2\Delta_c+2g+\chi b)}{\sqrt{2N}(\alpha-i\beta)},\nonumber\\
a_{-}=&\frac{\lambda J_{-}(2i\gamma+2\Delta_c-2g+\chi b^{\dagger})}{\sqrt{2N}(\alpha-i\beta)},
\label{super}
\end{align}
and
\begin{align}
p=&\frac{\lambda}{2i\gamma-(2g-\omega_m)}\left(\frac{\chi b}{2} + \frac{NU_0}{4} \right) \delta n \nonumber \\
&+\frac{\sqrt{2}\lambda}{\sqrt{N}[2i\gamma-(2g-\omega_m)]}(a_{+}+a^{\dagger}_{-})J_{-},
\label{inversion}
\end{align}
where
\begin{align}
\alpha=&\gamma^2+g^2-\Delta^2_c+\frac{NU_0}{4}\left[2\Delta_c+\chi\mathrm{Re}(b)\right]+\frac{\chi^2}{4}n_b,\nonumber\\
\beta=&\gamma\left(\Delta_c+\Delta^{\prime}_c\right),\nonumber
\end{align}
with $n_b=b^\dag b$. Substituting Eqs.\,(\ref{super}) and (\ref{inversion}) into the equation of the mechanical mode results in
\begin{equation}
\dot{b}=(-i\omega_m+i\omega^{\prime}+G-\gamma_m)b+C,
\end{equation}
where
\begin{align}
\omega^{\prime}=&\frac{\chi^2}{4\gamma^2+(2g-\omega_m)^2}\left[
\frac{(\omega_m-2g)\delta n}{4}
-\frac{2\gamma\lambda^2 J^2_{-}\beta}{N(\alpha^2+\beta^2)}
\right],\nonumber\\
C=&\frac{2\chi\lambda^2 J^2_{-}}{2\gamma+i(2g-\omega_m)}\left[\frac{NU_0\delta n}{16\lambda^2 J^2_{-}}-\frac{g\alpha+i(\alpha\gamma+\beta\Delta)}{N(\alpha^2+\beta^2)}\right],\nonumber
\end{align}
and the mechanical gain is $G=G_0+G_1$, with
\begin{align}
G_0=&\frac{\chi^2\gamma\delta n}{2(2g-\omega_m)^2+8\gamma^2},\nonumber\\
G_1=&-\frac{\chi^2\lambda^2J^2_{-}\beta(2g-\omega_m)}{N(\alpha^2+\beta^2)[(2g-\omega_m)^2+4\gamma^2]}.
\label{eq:gain}
\end{align}
$G_0$ is proportional to $\delta n$, which is in accordance with the results of Ref.\,\cite{Vahala2010}. Here, the population inversion is induced by the transverse pump field applied on the atomic gas, which has not been studied before.

\begin{figure}[ht]
\centering
\includegraphics[width=7.0cm]{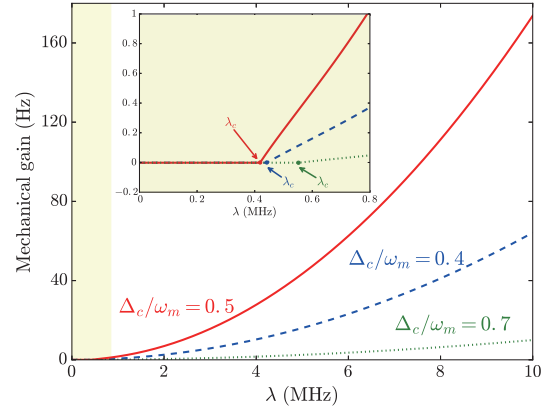}
\caption{The mechanical gain $G$ as a function of the atom-photon coupling $\lambda$. The inset shows an enlarged view of the mechanical gain for weak values of $\lambda$.}
\label{fig2}
\end{figure}

Figure\,\ref{fig2} shows the mechanical gain as a function of the atom-photon coupling.
We use experimentally accessible parameters in our calculations \cite{Wilson2009,Esslinger2010}, i.e, $N=10^5$, $\lambda_p=784.3\,$nm, $\gamma=2\pi\times 1\,$MHz, $\omega_m=2\pi\times 20\,$MHz, $\gamma_m=100\,$Hz, $g_0=2\pi\times 14\,$MHz, $NU_0=-2\pi\times2\,$MHz, $\chi=300\,$Hz, and $g=2\pi\times 10\,$MHz.
For a weak pump field ($\lambda<\lambda_c$), the atomic gas is in a normal phase, in which the steady-state photon number is safely neglected.
By increasing the pump power, the atomic gas enters into the superradiant phase for $\lambda>\lambda_c$.
We also find that the critical point $\lambda_c$ depends on the pump-cavity detuning $\Delta_c$, which can be minimized as $0.42\,$MHz at $\Delta_c/\omega_m\approx 0.5$. From Eqs.\,(\ref{eq:steady}) and (\ref{momentum}), the steady-state photon number for the second cavity is
$$
|a_{2,s}|^2=\frac{N\lambda^2(\Delta^2_c+\gamma^2)}{u^2+v^2}\left(1-\frac{\lambda^4_c}{\lambda^4} \right),
$$
which can further enhanced for a stronger pump. The photons in the left cavity also increases due to the optical tunneling, leading to an enhanced radiation pressure on the membrane. In this way, the membrane can be coherently driven by the superradiance. For simplicity, we only consider the phonon lasing near the threshold, so that the average phonon number $n_b$ is negligible and has negligible influence on the mechanical gain \cite{Wang2014}.
In the superradiance regime, the population inversion can be written as
\begin{equation}
\delta n = a^\dagger_{+}a_{+}-a^\dagger_{-}a_{-}\approx\frac{2g\Delta_c}{\Delta^2_c + \gamma^2}|a_{2,s}|^2.
\end{equation}
In addition, we use $2g=\omega_m$ in our calculations, so that the frequency difference of the optical supermodes equals the frequency of the mechanical mode. Therefore, the optical supermodes resonantly interact with the mechanical mode, which is beneficial for the phonon lasing \cite{Vahala2010}. Then we can calculate $G$ with the expression of $\delta n$, as shown in Fig.\,\ref{fig2}. Using the threshold condition $G=\gamma_{m}$ \cite{Vahala2010}, we can obtain the threshold atom-photon coupling $\lambda_{\mathrm{th}}$.
For example, we have $\lambda_{\rm th}=7.6\,$MHz with $\Delta_c/\omega_m=0.5$.

\begin{figure}[ht]
\centering             
\includegraphics[width=7.0cm]{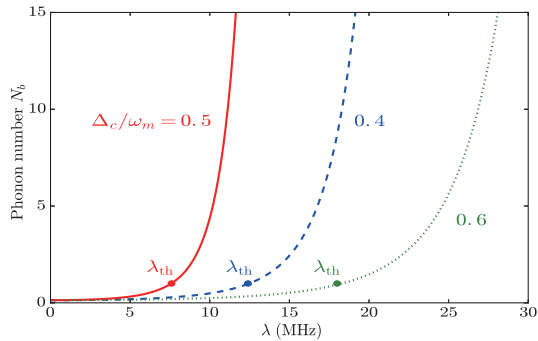}
\caption{Stimulated emitted phonon number $N_b$ as a function of the atom-photon coupling $\lambda$.}
\label{fig3}
\end{figure}

Figure\,\ref{fig3} shows the number of emitted phonons \cite{Jing2014,defect} $N_b = \exp[2(G - \gamma_m)/\gamma_m ]$ as a function of the atom-photon coupling.
In consideration of the peak intensity of the pump light $I=\epsilon_0 c E^2_0/2$, we can calculate the pump power as
\begin{equation}
P=\frac{1}{2}\pi w^2_cI=\frac{\pi \hbar^2 \epsilon_0 c w^2_c g^2_0\lambda^2}{2N\mathcal{D}^2U^2_0},
\end{equation}
where $w_c$ is the waist radius of the pump beam.
For $\Delta_c/\omega_m=0.5$, we have $P_{\mathrm{th}}\approx 6.4\,$mW, corresponding to $\lambda_{\rm th}=7.6\,$MHz, in which we use $w_c=25\,\mu$m \cite{Esslinger2010}.
We note that $N_b$ strongly depends on $\Delta_c$, and $\Delta_c/\omega_m=0.5$ is still the optimized condition for the phonon laser (see Fig.\,\ref{fig3}).

In conclusion, we have studied a superradiance-driven phonon laser in coupled cavities.
The atoms are driven by a transverse pump field, leading to superradiant light emissions. The enhanced coherent light can drive the membrane in the COM cavity, resulting in an enhanced phonon laser. Our work combines the COM-based phonon laser and the BEC-based superradiance, which provides a new way to generate and control phonon lasers. In the future, we will study the impact of the symmetry breaking at the superradiant phase transition on the behavior of the phonon lasing \cite{Esslinger2011}, superradiance-driven mechanical cooling, and mechanical squeezing with the assistance of atoms.

H.J. is supported by the NSF of China under Grants Nos. 11474087 and 11774086, and by the HuNU Talented Youth Foundation. We acknowledge Yu Chen at Capital Normal University for insightful discussions.

\end{document}